\documentclass[11pt,twoside]{atmp}

\usepackage{amsmath,amssymb}

\usepackage[all]{xy}
\usepackage{color}

\usepackage{latexsym}
\usepackage{amsfonts}
\usepackage{amssymb}
\usepackage{amsmath}
\usepackage{youngtab}

\def\cal{\mathcal}
\def\bydef{\equiv}

\def\R{\mathbb R}

\def\C{\mathbb C}
\def\be{\begin{equation}}
\def\ee{\end{equation}}
\def\bea{\begin{eqnarray}}
\def\eea{\end{eqnarray}}
\def\beal{\begin{align*}}
\def\eeal{\end{align*}}
\def\boxbox{{\text{\tiny \yng(2)}}}
\def\boxabox{{\text{\tiny \yng(1,1)}}}
\newcommand{\ket}[1]{\left|#1\right\rangle} 
\begin{document}

\title[Generalization of the Gell-Mann formula]
{Generalization of the Gell-Mann formula for $sl(n,\R)$ and
$su(n)$ algebras}


\author[I. Salom, Dj. \v Sija\v cki]{Igor Salom and Djordje \v Sija\v cki}

\address{Institute of Physics, University of Belgrade, P.O. Box 57, 11001
Belgrade, Serbia}  
\addressemail{isalom@ipb.ac.rs, sijacki@ipb.ac.rs}

\begin{abstract}
The so called Gell-Mann or decontraction formula is proposed as an
algebraic expression inverse to the In\"on\"u-Wigner Lie algebra
contraction. It is tailored to express the Lie algebra elements in
terms of the corresponding contracted ones. In the case of
$sl(n,\R)$ and $su(n)$ algebras, contracted w.r.t. $so(n)$
subalgebras, this formula is generally not valid, and applies only
in the cases of some algebra representations. A generalization of
the Gell-Mann formula for $sl(n,\R)$ and $su(n)$ algebras, that is
valid for all tensorial, spinorial, (non)unitary representations,
is obtained in a group manifold framework of the $SO(n)$ and/or
$Spin(n)$ group. The generalized formula is simple, concise and of
ample application potentiality. The matrix elements of the
$\overline{SL}(n,\R)/Spin(n)$, i.e. $SU(n)/SO(n)$, generators are
determined, by making use of the generalized formula, in a closed
form for all irreducible representations.
\end{abstract}

\maketitle

\section{Introduction}

Relations between various algebras/groups as well as their
subalgebras/sub\-groups played an important role in the Lie
algebras/groups theory and its representation theory development.
One of these relations is the well known In\"on\"u-Wigner
contraction \cite{InonuWigner} that was, besides its mathematical
merits, an important tool in numerous physical applications. There
is a variety of In\"on\"u-Wigner Lie algebra contraction
applications arising in various parts of Theoretical Physics. Just
to mention a few ranging from contractions from the Poincar\'e
algebra to the Galilean one, and from the Heisenberg algebras to
the Abelian ones of the same dimensions (a symmetry background of
a transition processes from relativistic and quantum mechanics to
classical mechanics) to those of contractions in the Kaluza-Klein
gauge theories framework, say from (Anti-)deSitter to the
Poincar\'e algebra, and various cases involving the Virasoro and
Kac-Moody algebras. For instance, our recent study of the Affine
Gauge Gravity Theory in $5D$ \cite{R13a} is heavily related to the
$sl(5,\R)$ algebra contraction w.r.t. its $so(1,3)$ subalgebra,
and the representations of the relevant algebras. In physical
terms, the meaning of the In\'on\'u-Wigner contraction is to
relate, for instance, strong to weak coupling regimes of the
corresponding theories, or geometrically curved to ``less curved'
and/or flat spaces.

It is not so well known that there is an inverse to the
In\"on\"u-Wigner contraction called the Gell-Mann formula
\cite{EncyclMath, HermannBook, Hermann, Berendt}. This formula is
a simple prescription designed to determine a deformation of a Lie
algebra that is ``inverse''to the In\"on\"u-Wigner contraction.
This formula relates elements of the starting algebra to the
corresponding ones of the contracted algebra. Moreover, by its
construction it relates also the representations of these two
algebras. Since, by a rule, various properties of the contracted
algebra are much easier to explore (e.g. construction of
representations \cite{Mackey}, decompositions of a direct product
of representations \cite{HermannBook}, etc.), this formula found
its place, as a useful and simple tool, even in some textbooks and
in the mathematical encyclopedia \cite{EncyclMath}.

To emphasize its meaning, this formula is also referred to as the
``decontraction'' formula. This formula was introduced by Dothan
and Ne'eman (while working at Caltech on noncompact algebras and
their representations) \cite{Dothan-Neeman} and advocated by
Hermann who learned from Gell-Mann about this formula, named it
after Gell-Mann, and made important contributions to the formula
himself.

\cutpage 

\setcounter{page}{2}

\noindent

In the general case the Gell-Mann formula construction goes as
follows: Let $\cal A$ be a symmetric Lie algebra $\cal A = M + T$
with a subalgebra
$\cal M$ such that:%
\be \cal [M, M]\subset M,\quad [M, T] \subset T,\quad [T, T]
\subset M.
                                                   \label{starting_algebra}
\ee %
Further, let $\cal A'$ be its In\"on\"u-Wigner contraction algebra
w.r.t its
subalgebra $\cal M$, i.e. $\cal A' = M + U$, where %
\be {\cal [M, M]\subset M,\quad [M, U] \subset U,\quad [U, U]} =
\{0\}.
\ee %
The Gell-Mann formula states that the elements $T \in {\cal T}$
can be in certain cases expressed in terms of the contracted
algebra elements $M \in
{\cal M}$ and $U \in {\cal U}$ by the following rather simple expression:%
\be T = i \frac{\alpha}{\sqrt{U\cdot U}}[C_2({\cal M}), U] +
\sigma U.
                                                   \label{GM_general}
\ee %
Here, $C_2({\cal M})$ and $U\cdot U$ denote the second order
Casimir operators of the $\cal M$ and $\cal A'$ algebras
respectively, while $\alpha$ is a normalization constant and
$\sigma$ is an arbitrary parameter. For a mathematically strict
definition, cf. \cite{EncyclMath}.

The main drawback of the Gell-Mann formula is its limited
validity. There is a number of references dealing with the
question when this formula is applicable \cite{HermannBook,
Hermann, Berendt, Salom+Sijacki_validity}. The formula is best
studied in the case of (pseudo) orthogonal algebras $so(m,n)$
contracted w.r.t.  their $so(m-1,n)$ and/or $so(m,n-1)$
subalgebras, i.e. in the corresponding group cases: $SO(m,n)
\rightarrow R^{m+n-1} \wedge SO(m-1,n)$ and/or $SO(m,n)
\rightarrow R^{m+n-1} \wedge SO(m,n-1)$, where, loosely speaking,
the Gell-Mann formula works very well \cite{Sankara}. Moreover,
the case of (pseudo) orthogonal algebras is the only one where
this formula is valid for (almost) all representations
\cite{Weimar}. Recently, we studied the $sl(n,\R)$ cases where the
Gel-Mann formula does not hold as a general operator expression
and its validity depends heavily on the $sl(n,\R)$ algebra
representation space. An exhaustive list of the cases for which
the Gell-Mann formula for $sl(n,\R)$ algebras hold is obtained
\cite{Salom+Sijacki_validity}.

There were some attempts to generalize the Gell-Mann formula for
the ``decontracted''algebra operators of the complex simple Lie
algebras $g$ with respect to decomposition $g = k + i k = k_c$
\cite{Stovicek, Mukunda}, that resulted in a form of relatively
complicated polynomial expressions.

In this work we consider Gell-Mann's formula in the $sl(n,\R)$
algebra cases, where the contraction is performed w.r.t. their
maximal compact $so(n)$ subalgebras. The Gell-Mann formula in
these cases is especially  valuable as a tool in the problem of
finding all unitary irreducible representations of the $sl(n,\R)$
algebras in the basis of the $SO(n)$ and/or $Spin(n)$ groups
generated by their $so(n)$ subalgebras. Finding representations in
the basis of the maximal compact subgroup $SO(n)$ of the
$SL(n,\R)$ group, is mathematically superior, and it suites well
various physical applications in particular in nuclear physics,
gravity, physics of p-branes \cite{SijackiBranes} etc. As an
example consider a gauge theory based on the Affine spacetime
symmetry $SA(n,\R)$ $=$ $T_{n}\wedge \overline{SL}(n,\R)$. The
gauge covariant derivative, $D_{\mu}$, $\mu = 0, 1, \dots , n-1$,
as acting on an Affine matter field $\Psi (x)$, is given by,
$$
D_{\mu} \Psi (x)_i = \left( \partial_{\mu} -i \Gamma^{ab}_{\mu}
(x) \left(
  Q_{ab}\right)_i^j \right) \Psi (x)_j ,\quad Q_{ab} \in sl(n,\R) ,
$$
where $\Gamma^{ab}_{\mu} (x)$ are the $sl(n,\R)$ connections, and
$i,j$ enumerate the matter field components. The matter--gravity
vertices require the knowledge of the $sl(n,\R)$ operators matrix
elements $\left( Q_{ab}\right)_i^j$ in the Hilbert space of the
matter field components $\{ \Psi_i (x) \}$. In particular, for a
generic spinorial $\overline{SL}(n,\R)$ matter field, en explicit
form of the matrix elements of the $sl(n,\R)$ generators for
infinite-dimensional representation corresponding to the $\Psi$
field is required.

Moreover, this framework opens up a possibility of finding, in a
rather straightforward manner, all matrix elements of noncompact
$SL(n,\R)$, i.e. $\overline{SL}(n,\R)$ generators for all finite
and infinite dimensional representations. Unfortunately, as stated
above, the original Gell-Mann formula is of limited validity, and
it can be applied in the classes of multiplicity free
representations only. Recently we have demonstrated, by an
explicit construction, that the Gell-Mann formula has a
generalization that is valid for all irreducible representations
of the $sl(n,\R)$, $n = 3,4$ and $5$ algebras
\cite{Salom-Sijacki-IJGMMP}. These formulas, though comparatively
simpler than the ones resulting from some other attempts to
generalize the Gell-Mann formula, are still rather cumbersome
taking a few rows. In this paper we present a generalization of
the Gell-Mann formula that is almost as compact and simple as the
original formula, which is valid for \emph{ all
  representations of the $sl(n,\R)$ algebras for all $n$ values}.

In our previous study of the Gell-Mann formula and its
generalization for the $sl(n,\R)$, $n=3,4,5$ algebras
\cite{Salom-Sijacki-IJGMMP}, we first extracted the generalized
formula from the previously known generic expressions in the
$sl(3,\R)$ and $sl(4,\R)$ cases which enabled us to obtain finally
the generalized formula for the $sl(5,\R)$ algebra. A crucial
ingredient in generalization of the Gell-Mann formula for the
$sl(n,\R)$ algebras is a role of the ``left-rotation'' action of
the $Spin(n)$ subgroup of the corresponding $\overline{SL}(n,\R)$
group, which manages the non trivial multiplicity of the maximal
compact $so(n)$ subalgebra representations for a generic
$sl(n,\R)$ representation. In this work we show, by making use of
the Cartesian basis, how to obtain the explicit form of the
generalized Gell-Mann formula for an arbitrary $sl(n,\R)$ algebra
with a left-rotations included properly and demonstrate the
closure of the $sl(n,\R)$ algebra commutation relations. Moreover,
we showed how to rewrite the expression of the generalized formula
in a suitable basis (e.g. the Gel'fand-Tsetlin basis) allowing us
to directly write down matrix expressions of the $sl(n,\R)$, i.e.
$su(n)$, generators for an arbitrary irreducible representation in
the basis of the $Spin(n)$ group.

Note that due to mutual relations between the $sl(n,\R)$ and
$su(n)$ algebras, one can convey the Gell-Mann formula results
obtained for the $sl(n,\R)$ algebras to the corresponding ones of
the $su(n)$ algebras. Though, there are some subtleties in that
process that are considered below.

\section{$sl(n,\R)$ and $SL(n,\R)$ topology and representations
  considerations}

As already stated, the Gell-Mann formula, except in the cases of
(pseudo) orthogonal algebras, is not generally valid by itself,
and its validity depends on the representations of the algebra as
well. Therefore, in the case of the $SL(n,\R)$ groups, i.e.\ their
$sl(n,\R)$ algebras, one faces, in addition to the pure algebraic
features, the matters that are relevant to the group/algebra
representations theory as well: notably (i) the  group topology
properties, and (ii) the non trivial multiplicity of the
$SL(n,\R)$, and $sl(n,\R)$ representations in the $SO(n)$, and
$so(n)$ basis, respectively. Both features are rather subtle for
$n \geq 3$. Note that, in the case of the $sl(n,\R)$ algebras, due
to a fact that the generalization of the Gell-Mann formula depends
on the algebra irreducible representation features, the
construction itself deviates from the standard Lie algebra
deformation approach.

The $SL(n,\R)$ group can be decomposed, as any semisimple Lie
group, into the product of its maximal compact subgroup $K =
SO(n)$, an Abelian group $A$ and a nilpotent group $N$. It is well
known that only $K$ is not guaranteed to be simply-connected.
There exists a universal covering group $\overline{K} =
\overline{SO}(n)$ of $K = SO(n)$, and thus also a universal
covering of $G =SL(n,\R)$: $\overline{SL}(n,\R) \simeq
\overline{SO}(n) \times A \times N$. For $n \geq 3$, $SL(n,\R)$
has double covering, defined by $\overline{SO}(n) \simeq Spin(n)$
the double-covering of the $SO(n)$ subgroup. The universal
covering group $\overline{G}$ of a given group $G$ is a group with
the same Lie algebra and with a simply-connected group manifold. A
finite dimensional covering, $\overline{SL}(n,\R)$ exists provided
one can embed $\overline{SL}(n,\R)$ into a group of finite complex
matrices that contain $Spin(n)$ as a subgroup. A scan of the
Cartan classical algebras points to the $SL(n,\C)$ groups as a
natural candidate for the $SL(n,\R)$ groups covering. However,
there is no match of the defining dimensionalities of the
$SL(n,\C)$ and $Spin(n)$ groups for $n \geq 3$, $dim(SL(n, \C)) =
n < 2^{\left[\frac{n-1}{2}\right] } = dim(Spin(n))$, except for $n
= 8$. In the $n = 8$ case, one finds that the orthogonal subgroup
of the $SL(8,\R)$ and $SL(8, \C)$ groups is $SO(8)$ and not
$Spin(8)$. For a detailed account of the $D=4$ case cf.
\cite{YN+DjS-IJMPA}. Thus, we conclude that there are no covering
groups of the $SL(n,\R)$, $n \geq 3$ groups defined in
finite-dimensional spaces. An explicit construction of all
$SL(3,R)$ irreducible representations, unitary and nonunitary
multiplicity-free spinorial \cite{DjS-JMP31}, and unitary
non-multiplicity-free \cite{SijackiSL3}, shows that they are
infinite-dimensional. The universal (double) covering groups,
$\overline{SL}(n,\R)$, $n \geq 3$ of the $SL(n,\R)$, $n \geq 3$
group are groups of infinite complex matrices. All their spinorial
representations are infinite dimensional. In the reduction of this
representations w.r.t. $Spin(n)$ subgroups, one finds $Spin(n)$
representations of unbounded spin values.

The $SU(n)$ groups are compact, with a simply connected group
manifold, thus being its own universal coverings. The $SO(n)$
subgroups are embedded into the $SU(n)$ groups as $n$-dimensional
matrices, and this embedding does not allow nontrivial (double)
covering of $SO(n)$ within $SU(n)$. As a consequence, in the
reduction of the $SU(n)$ unitary irreducible representations one
finds the tensorial $SO(n)$ representations only.

An inspection of the unitary irreducible representations of the
$\overline{SL}(n,\R)$, $n = 3,4$  groups \cite{SijackiSL3,
SijackiSL4} shows that they have, as a rule, a nontrivial
multiplicity of the $Spin(n)$, $n = 3, 4$ subgroup
representations. It is well known, already from the case of the
$SU(3)$ representations in the $SO(3)$ subgroup basis, that the
additional labels required to describe this nontrivial
multiplicity cannot be solely related to the group generators
themselves. An elegant solution, that provides the required
additional labels, is to work in the group manifold of the $SO(n)$
maximal compact subgroup, and to consider an action of the group
both to the right and to the left. In this way one obtains,
besides the maximal compact subgroup labels, an additional set of
labels to describe the $SO(n)$ subgroup multiplicity.

\section{In\"on\"u-Wigner contraction of $sl(n, \R)$ algebras}

The $sl(n,\R)$ algebra operators, i.e.\ the $SL(n,\R)$,
$\overline{SL}(n,\R)$ group generators, can be split into two
subsets: $M_{ab}$, $a,b = 1, 2, ...,n$ operators of the maximal
compact subalgebra $so(n)$ (corresponding to the antisymmetric
real $n\times n$ matrices, $M_{ab} = -M_{ba}$), and the, so
called, sheer operators $T_{ab}$, $a,b = 1, 2, ...,n$
(corresponding to the symmetric traceless real $n\times n$
matrices, $T_{ab} = T_{ba}$). The $sl(n,\R)$ commutation
relations, in this basis, read: \bea [M_{ab},M_{cd}] &=&
i(\delta_{ac}M_{bd} + \delta_{ad}M_{cb} - \delta_{bc}M_{ad} -
\delta_{bd}M_{ca}),
                                                 \label{MMcommutator1} \\
{}[M_{ab},T_{cd}] &=& i(\delta_{ac}T_{bd} + \delta_{ad}T_{cb} -
\delta_{bc}T_{ad} - \delta_{bd}T_{ca}),
                                                 \label{MTcommutator}\\
{}[T_{ab},T_{cd}] &=& i(\delta_{ac}M_{db} + \delta_{ad}M_{cb} +
\delta_{bc}M_{da} + \delta_{bd}M_{ca}).          \label{TTcommutator}\eea %

The $su(n)$ algebra operators can be split likewise w.r.t. its
$so(n)$ subalgebra into $M_{ab}$ and $T^{su(n)}_{ab}$, $a,b = 1,
2, ...,n$. The $T^{su(n)}_{ab}$ and $T_{ab}$ operators are
mutually related by $T^{su(n)}_{ab}$ = i $T_{ab}$, and the
$[T^{su(n)}_{ab}, T^{su(n)}_{cd}]$ differs from
(\ref{TTcommutator}) by having an overall plus sign on the
right-hand side.

The In\"on\"u-Wigner contraction of $sl(n, \R)$ with respect to
its
maximal compact subalgebra $so(n)$ is given by the limiting procedure: %
\be U_{ab} \bydef \lim_{\epsilon \rightarrow 0} (\epsilon T_{ab}),
                                                 \label{limit} \ee %
which leads to the following commutation relations: %
\bea [M_{ab},M_{cd}] &=& i(\delta_{ac}M_{bd} +
\delta_{ad}M_{cb} - \delta_{bc}M_{ad} - \delta_{bd}M_{ca}) \\
{}[M_{ab},U_{cd}] &=& i(\delta_{ac}U_{bd} + \delta_{ad}U_{cb} -
\delta_{bc}U_{ad} - \delta_{bd}U_{ca})
                                                  \label{MUcommutator} \\
{}[U_{ab},U_{cd}] &=& 0. \eea %

Therefore, the In\"on\"u-Wigner contraction of $sl(n, \R)$ gives a
semidirect sum  $r_{\frac{n(n+1)}{2}-1}\biguplus so(n)$ algebra,
where $r_{\frac{n(n+1)}{2}-1}$ is an Abelian subalgebra (ideal) of
``translations'' in $\frac{n(n+1)}{2}-1$ dimensions.

The Gell-Mann formula, a prescription to provide an ``inverse''to
the
In\"on\"u-Wigner contraction, (\ref{limit}), in this case reads: %
\be T^{\sigma}_{ab} = \frac{i\alpha}{\sqrt{U\cdot U}}
[C_2(so(n)), U_{ab}] + \sigma U_{ab} ,   \label{Gell-Mann_original}\ee %
where $C_2(so(n))$ denotes the second order Casimir operator of
the $so(n)$ subalgebra, $\frac 12 \sum M_{ab} M_{ab}$, while
$\sigma$ is an arbitrary (complex) parameter and $\alpha$ is a
(real) normalization constant that depends on $n$. This expression
can also be written
in an equivalent form: %
\be T^{\sigma'}_{ab} =  - \frac{2\alpha}{\sqrt{U\cdot U}} \sum_c
U_{c\{\!a}M_{b\}c} + \sigma' U_{ab},
                                         \label{Gell-Mann_original_v2}\ee %
where $\sigma'$ differs from $\sigma$ accordingly, and $\{\ \}$
denotes symmetrization of the enclosed indices.

In order to make use of the Gell-Mann formula to obtain the
$sl(n,\R)$ representations, the first necessary step is to
determine representation matrix elements of the contracted algebra
operators. The corresponding contracted  group is a semidirect
product of $SO(n)$ and an Abelian group, and it is well known that
the usual group induction method provides the complete set of all
inequivalent irreducible representations \cite{Mackey}.
Nevertheless, we will not pursue the induction approach here.
Instead, we will rather proceed to work in the representation
space of square integrable functions ${\cal L}^2(Spin(n))$ over
the $Spin(n)$ group (in accord with the $SL(n,\R)$ topological
properties), with the standard invariant Haar measure. As for our
final goal, this approach ensures certain advantages: (i) The
generalized Gell-Mann formula is expressed in terms of tensor
operators w.r.t. the maximal compact subgroup basis (instead
w.r.t. the eigenvector basis of the Abelian subgroup), (ii) This
representation space allows for all inequivalent irreducible
representations of the contracted group (some of the irreducible
representations are multiply contained, i.e. each such
representation appears as many times as is the dimension of the
corresponding little group representation and all of them,
irrespectively of the corresponding stabilizer, can be treated in
an unified manner), and (iii) this space is rich enough to contain
all representatives from equivalence classes of the
$\overline{SL}(n,\R)$ group, i.e.\ $sl(n,\R)$ algebra
representations \cite{HarishChandra}. The last feature provides
the necessary requirement of a framework needed for generalization
of the Gell-Mann formula, i.e. a unique framework providing for
all $sl(n,\R)$ (unitary) irreducible representations.

The generators of the contracted group are generically represented
in this space as follows. The $so(n)$ subalgebra operators act, in
a standard way, via
the group action to the right: %
$$
M_{ab} \ket{\phi} = -i \frac{d}{dt} \exp(i t M_{ab})\Big|_{t=0}
\ket{\phi}, \quad g' \ket{g} = \ket{g'g}, \quad \ket{\phi} \in
{\cal L}^2(Spin(n)).
$$
The Abelian operators $U_{ab}$ act multiplicatively as Wigner's
$D$-functions (the appropriate representation $SO(n)$ group matrix
elements as functions of
the group parameters): %
\be U_{ab} \rightarrow |u| D^{\boxbox}_{v (ab)}\!(g^{-1}) \equiv
\left<
\begin{array}{c} \boxbox \\ v \end{array} \right| g^{-1}
\left| \begin{array}{c} \boxbox \\ ab \end{array} \right> ,
\label{UisD}\ee %
$|u|$ being a constant norm, $g(\theta)$ being an $SO(n)$ element,
and in order to simplify notation we denote by $\boxbox$ (in a
parallel to the Young tableaux) the symmetric second rank tensor
representation of $SO(n)$. The vector $\left| \begin{array}{c}
\boxbox \\ ab \end{array} \right>$ from the $\boxbox$
representation space is determined by the $ab$ ``double'' index of
$U_{ab}$, whereas the vector $\left| \begin{array}{c} \boxbox \\ v
\end{array} \right>$ can be an arbitrary vector belonging to the $\frac 12
n(n+1)-1$ dimensional $\boxbox$ representation (the choice of $v$
is determined, in Wigner's terminology, by the little group of the
obtained representation). Taking an inverse of $g$ in (\ref{UisD})
ensures the correct transformation properties. The form of the
representation of the Abelian operators merely reflects the fact
that they transform as symmetric second rank tensor w.r.t $so(n)$
(\ref{MUcommutator}) and that they mutually commute.

A natural discrete orthonormal basis in the $Spin(n)$
representation space is given by properly normalized functions of
the $Spin(n)$ representation matrix
elements: %
\bea &&\left\{\left| {\begin{array}{l@{}l} \{J\} & \\ \{k\} &
\{m\} \end{array}}
  \right>  \equiv \int {\scriptstyle \sqrt{dim(\{J\})}}
D^{\{J\}}_{\{k\}\{m\}}\!(g(\theta)^{-1}) d\theta
\ket{g(\theta)}\right\},
                                                   \label{naturalbasis}  \\
&&\left< { \begin{array}{l@{}l|l@{}l}  \{J'\} & & \{J\}\\ \{k'\} &
\{m'\} & \{k\} & \{m\} \end{array}} \right> =
\delta_{\{J'\}\{J\}}\delta_{\{k'\}\{k\}}\delta_{\{m'\}\{m\}},
\nonumber
                                                  \eea
where $d\theta$ is an (normalized) invariant Haar measure,
$D^{\{J\}}_{\{k\}\{m\}}$ are the $Spin(n)$ irreducible
representation matrix
elements, %
\be D^{\{J\}}_{\{k\}\{m\}}(\theta) \equiv \left< \begin{array}{c}
\{J\} \\\{k\} \end{array} \right| R(\theta) \left|
\begin{array}{c} \{J\} \\\{m\} \end{array} \right>.
\ee %
Here, $\{J\}$ stands for a set of the $Spin(n)$ irreducible
representation labels, while $\{k\}$ and $\{m\}$ labels enumerate
the $dim(D^{\{J\}})$ representation basis vectors.

An action of the $so(n)$ operators in this basis is well known,
and it can be written in terms of the Clebsch-Gordan coefficients
of the $Spin(n)$ group as follows,
{\renewcommand{\arraystretch}{0.2} %
\be \left< { \begin{array}{l@{}l} \{J'\} & \\ \{k'\} & \{m'\}
\end{array}} \right| M_{ab} \left| {\begin{array}{l@{}l} \{J\} &
\\ \{k\} & \{m\} \end{array}} \right> = \delta_{\{J'\}\{J\}}
{\scriptstyle \sqrt{C_2(\{J\})}} \;
C\!\!\!{\scriptsize\begin{array}{c@{}c@{}c} \{J\} & \boxabox & \{J'\} \\
\{m\} & (ab) & \{m'\} \end{array}}.              \label{Maction}\ee} %
The matrix elements of the $U_{ab}$ operators in this basis are
readily found
to read: %
{\renewcommand{\arraystretch}{0.2} \bea &&\left<{
\begin{array}{l@{}l} \{J'\} & \\ \{k'\} & \{m'\}\end{array}}
\right| U^{v}_{ab} \left| {\begin{array}{l@{}l} \{J\} & \\ \{k\} &
\{m\} \end{array}}\right>
\nonumber \\ %
&&= |u|\left< { \begin{array}{l@{}l} \{J'\} & \\ \{k'\} & \{m'\}
\end{array}} \right| D^{-1\boxbox}_{v (ab)} \left|
{\begin{array}{l@{}l} \{J\} & \\ \{k\} & \{m\} \end{array}}
\right>
\nonumber \\ %
&&= |u|{\scriptstyle \sqrt{dim(\{J'\})dim(\{J\})}} \int
D_{\{k'\}\{m'\}}^{\{J'\}*}\!(\theta) D^\boxbox_{v (ab)}(\theta)
D_{\{k\}\{m\}}^{\{J\}}(\theta) d\theta
                                                      \label{Uaction} \\
&&= |u|{\scriptstyle \sqrt{\frac{dim(\{J\})}{dim(\{J'\})}}}
C\!\!{\scriptsize
\begin{array}{c@{}c@{}c} \{J\} & \boxbox & \{J'\} \\ \{k\} & v & \{k'\}
\end{array}} C\!\!{\scriptsize
\begin{array}{c@{}c@{}c} \{J\} & \boxbox & \{J'\} \\ \{m\} & (ab) & \{m'\}
\end{array}} . \nonumber \eea} %
A closed form of the matrix elements of the whole contracted
algebra $r_{\frac{n(n+1)}{2}-1}\biguplus so(n)$ representations is
thus explicitly given in this space by (\ref{Maction}) and
(\ref{Uaction}).

\section{\label{sec:GeneralizedGM}The generalized formula}

The Gell-Mann formula (\ref{Gell-Mann_original}) expressed in
terms of the
Wigner's functions now reads: %
\be T^{\sigma}_{ab} = i\alpha [C_2(so(n)), D^\boxbox_{v (ab)}] +
\sigma
D^\boxbox_{v (ab)}.             \label{Gell-Mann_original in D terms}\ee %
However, the $T_{ab}$ operators, as given by this expression
depending on the representation space, do not close upon the
$sl(n,\R)$ commutation relations (\ref{TTcommutator}) in the whole
representation space. Indeed, we have shown in
\cite{Salom+Sijacki_validity} that this formula yields the
$sl(n,\R) / so(n)$ generators only when the representation space
is restricted to $Spin(n)/(Spin(m)\otimes Spin(n-m))$, with
$\left| \begin{array}{c} \boxbox \\ v \end{array} \right>$ vector
chosen to be invariant with respect to the $Spin(m)\otimes
Spin(n-m)$, $1\leq m\leq n-1$, while $Spin(1)$ is here implied to
be a trivial group.

We have shown in \cite{Salom-Sijacki-IJGMMP} that the Gell-Mann
formula can be generalized in the cases of the $sl(n,\R)$, $n=3,4$
and $5$ algebras so that the commutation relations
(\ref{TTcommutator}) hold for an arbitrary irreducible
representation in whole Hilbert space over the $Spin(n)$ group.
This generalization was achieved by adding certain extra terms
containing the generators of $SO(n)$ subgroup action to the left.
For example, the generalized formula in the $sl(5,\R)$ case reads:
{\renewcommand{\arraystretch}{0.2} %
\be\begin{array}{rl} %
T^{\sigma_1 \sigma_2 \delta_1 \delta_2}\!\!{\tiny
\begin{array}{l@{\!}l} \vphantom {\overline 1} &
\\ j_1 & j_2 \\
\vphantom{\underline{\mu_1}}\mu_1 & \mu_2
\end{array}} =&
\sigma_1 D\!\!{\tiny
\begin{array}{l@{}l@{}l@{}l} \overline 1 &
\overline 1 & & \\ 0 & 0 & j_1 & j_2 \\
\vphantom{\underline{m_1}}0 & 0 &\mu_1 & \mu_2
\end{array}} +
i\sqrt{\frac 15}[C_2(so(5)), D\!\!{\tiny
\begin{array}{l@{}l@{}l@{}l} \overline 1 &
\overline 1 & & \\ 0 & 0 & j_1 & j_2 \\
\vphantom{\underline{m_1}}0 & 0 &\mu_1 & \mu_2
\end{array}}]
\\ &
 + i\Bigg(\sigma_2 D\!\!{\tiny
\begin{array}{l@{}l@{}l@{}l} \overline 1 &
\overline 1 & & \\ 1 & 1 & j_1 & j_2 \\
\vphantom{\underline{m_1}}0 & 0 &\mu_1 & \mu_2
\end{array}} +
{\textstyle \frac 12}[C_2(so(4)_K), D\!\!{\tiny
\begin{array}{l@{}l@{}l@{}l} \overline 1 &
\overline 1 & & \\ 1 & 1 & j_1 & j_2 \\
\vphantom{\underline{m_1}}0 & 0 &\mu_1 & \mu_2
\end{array}}]
\\ &
 - D\!\!{\tiny
\begin{array}{c@{}c@{}l@{}l} \overline 1 &
\overline 1 & & \\ 1 & 1 & j_1 & j_2 \\
\vphantom{\underline{m_1}}1 & -1 &\mu_1 & \mu_2
\end{array}} (\delta_1 + K\!\!{\tiny
\begin{array}{l@{}l} \overline 1 &
\overline 0 \\ 1 & 0 \\
\vphantom{\underline{m_1}}0 & 0
\end{array}} - K\!\!{\tiny
\begin{array}{l@{}l} \overline 1 &
\overline 0 \\ 0 & 1 \\
\vphantom{\underline{m_1}}0 & 0
\end{array}})-
D\!\!{\tiny
\begin{array}{c@{}c@{}l@{}l} \overline 1 &
\overline 1 & & \\ 1 & 1 & j_1 & j_2 \\
\vphantom{\underline{m_1}}-1 & 1 &\mu_1 & \mu_2
\end{array}} (\delta_1 - K\!\!{\tiny
\begin{array}{l@{}l} \overline 1 &
\overline 0 \\ 1 & 0 \\
\vphantom{\underline{m_1}}0 & 0
\end{array}} + K\!\!{\tiny
\begin{array}{l@{}l} \overline 1 &
\overline 0 \\ 0 & 1 \\
\vphantom{\underline{m_1}}0 & 0
\end{array}})%
\\ &
 + D\!\!{\tiny
\begin{array}{c@{}c@{}l@{}l} \overline 1 &
\overline 1 & & \\ 1 & 1 & j_1 & j_2 \\
\vphantom{\underline{m_1}}1 & 1 &\mu_1 & \mu_2
\end{array}} (\delta_2 + K\!\!{\tiny
\begin{array}{l@{}l} \overline 1 &
\overline 0 \\ 1 & 0 \\
\vphantom{\underline{m_1}}0 & 0
\end{array}} + K\!\!{\tiny
\begin{array}{l@{}l} \overline 1 &
\overline 0 \\ 0 & 1 \\
\vphantom{\underline{m_1}}0 & 0
\end{array}})+
D\!\!{\tiny
\begin{array}{c@{}c@{}l@{}l} \overline 1 &
\overline 1 & & \\ 1 & 1 & j_1 & j_2 \\
\vphantom{\underline{m_1}}-1 & -1 &\mu_1 & \mu_2
\end{array}} (\delta_2 - K\!\!{\tiny
\begin{array}{l@{}l} \overline 1 &
\overline 0 \\ 1 & 0 \\
\vphantom{\underline{m_1}}0 & 0
\end{array}} - K\!\!{\tiny
\begin{array}{l@{}l} \overline 1 &
\overline 0 \\ 0 & 1 \\
\vphantom{\underline{m_1}}0 & 0
\end{array}})\Bigg)
\end{array} ,                                       \label{sl5GGM}\ee} %

The left action generators $K$, that appear in the formula can be
related to
the $M_{ab}$ operators by the following expression: %
\be K_{ab} \equiv
g^{(a''b'')(a'b')} D_{(ab)(a''b'')}^\boxabox M_{a'b'}, \label{Koperator}\ee %
where $g^{(a''b'')(a'b')}$ is the Cartan metric tensor of $SO(n)$.
The $K_{ab}$ operators behave exactly as the rotation generators
$M_{ab}$, it is only that they act on the lower left-hand side
indices of the basis (\ref{naturalbasis}):
{\renewcommand{\arraystretch}{0.2} %
\be \left< { \begin{array}{l@{}l} \{J'\} & \\ \{k'\} & \{m'\}
\end{array}} \right| K_{ab} \left| {\begin{array}{l@{}l} \{J\} &
\\ \{k\} & \{m\} \end{array}} \right> = \delta_{\{J'\}\{J\}}
{\scriptstyle \sqrt{C_2(\{J\})}} \; C\!\!\!{\scriptsize
\begin{array}{c@{}c@{}c} \{J\} & \boxabox & \{J'\} \\ \{k\} & (ab) & \{k'\}
\end{array}}.                                          \label{Kaction}\ee} %
Due to the fact that the mutually contragradient $SO(n)$
representations are equivalent, the $K_{ab}$ operators are
directly related to the "left" action of the $SO(n)$ subgroup on
${\cal L}^2(\ket{g(\theta)})$: $g' \ket{g} = \ket{g {g'}^{-1}}$.
The $K_{ab}$ and $M_{ab}$ operators mutually commute, however, the
corresponding Casimir operators match, i.e.\ $K_{ab}^2 =
M_{ab}^2$.

The generalized Gell-Mann formulas for $sl(3, \R)$, $sl(4, \R)$
and $sl(5,\R)$ \cite{Salom-Sijacki-IJGMMP} are given by rather
cumbersome expressions. However, when these formulas are expressed
in the Cartesian basis (like formulas
(\ref{MMcommutator1})-(\ref{TTcommutator})) in terms of the
$K_{ab}$ operators and anti-commutators rather than commutators
the resulting expressions become extremely simple. Moreover, this
form allows for an immediate generalization to the case of an
arbitrary $n$. We prove below that the generalized Gell-Mann
formula for any
$sl(n,\R)$ algebra w.r.t its $so(n)$ subalgebra takes the following form:%
\be T^{\sigma_2 \dots \sigma_n}_{ab} = i\sum^n_{c>d} \{ K_{cd},
D^\boxbox_{(cd)(ab)} \} +
i\sum_{c=2}^n \sigma_c D^\boxbox_{(cc)(ab)},             \label{GGM} \ee %
where $\sigma_c$ is a set of $n-1$ arbitrary parameters that
essentially (up to some discrete parameters) label $sl(n,\R)$
irreducible representations. Note that the sum in the first term
goes only over pairs $(c,d)$ where $c>d$ i.e.\ it is not symmetric
in $c,\ d$.

Let us begin the proof that the expressions (\ref{GGM}) satisfy
the $sl(n,\R)$
commutation relation (\ref{TTcommutator}) by introducing operators: %
\be T_{ab}^{[c]} = i\sum_{d = 1}^{c-1} \{ K_{cd},
D^\boxbox_{(cd)(ab)} \}
+ i\sigma_c D^\boxbox_{(cc)(ab)},\quad c=2,\dots ,n   \label{Tc operators}\ee %
i.e. expressing the generalized expression (\ref{GGM}) as: %
\be T_{ab} = \sum_{c=2}^{n} T_{ab}^{[c]}.
\ee %
A straightforward, but somewhat lengthy, calculation yields
$[T_{ab}^{(c)}, T_{a'b'}^{(d)}]=[T_{a'b'}^{[c]}, T_{ab}^{[d]}]$
for $c\neq d$,
and thus we find: %
\be[T_{ab}, T_{a'b'}] =\sum_{c}[T_{ab}^{[c]}, T_{a'b'}^{[c]}]=
-i\sum_{c,d,d'} \{K_{dd'}, \{D^\boxbox_{(cd)(ab)},
D^\boxbox_{(cd')(a'b')}\}\}. \label{calculation1}\ee %
By making use of the identity: %
\bea & \sum_{c} (D^\boxbox_{(cd)(ab)} D^\boxbox_{(cd')(a'b')} -
D^\boxbox_{(cd')(ab)} D^\boxbox_{(cd)(a'b')}) \\
& = \frac 12(\delta_{aa'} D^\boxabox_{(dd')(bb')} + \delta_{bb'}
D^\boxabox_{(dd')(aa')} + \delta_{ab'} D^\boxabox_{(dd')(ba')} +
\delta_{ba'} D^\boxabox_{(dd')(ab')}) \nonumber \eea %
and the fact that the $M$ generators are given in terms of the $K$
operators via the $D^\boxabox$ operators (cf. (\ref{Koperator})),
one verifies the desired expression (\ref{TTcommutator}).

Note that the first equality in (\ref{calculation1}) implies that
the overall sign of operators $T_{ab}^{[c]}$ is inessential.
Moreover, any left rotation (generated by the $K$ operators) of
the generalized formula (\ref{GGM}) will preserve the $[T, T]$
commutator (\ref{TTcommutator}) and thus lead to another valid
expression for the generalized Gell-Mann formula. The generalized
formulas related in this way form an equivalence class of formulas
that yield the same set of $sl(n,\R)$ irreducible representations.
Besides this class there are a few alternative useful expressions
of the generalized Gell-Mann formula. We point out explicitly two
cases below.

Let us consider operators: %
\be U^{(cd)}_{ab} \equiv D^\boxbox_{(cd)(ab)} \ee %
stressing that $D^\boxbox_{(cd)(ab)}$ is just a particular
representation of the $U_{ab}$ operators (\ref{UisD}),
characterized by the choice of the vector $v$ to be $v = (cd)$ and
$|u| = 1$. Then, by making use of the commutation
relations to shift the $K$ operators to the right in (\ref{GGM}) we find : %
\be T_{ab} = 2i \sum^n_{c>d} U^{(cd)}_{ab} K_{cd} +
i\sum_{c=2}^n \sigma'_c U^{(cc)}_{ab}.                   \label{GGMv2} \ee %
This expression for the generalized formula can now be directly
compared to the original formula in the form
(\ref{Gell-Mann_original_v2}). It is as simple as the original
Gell-Mann formula, with a crucial advantage of being valid in the
whole representation space over ${\cal L}^2(Spin(n))$. General
validity of the new formula is reflected in the fact that there
are now $n-1$ free parameters, i.e.\ representation labels,
matching the $sl(n,\R)$ algebra rank, compared to just one
parameter of the original Gell-Mann formula.

Another notable form of the generalized formula relies on the fact
that the
operators $T^{[c]}$ (\ref{Tc operators}) can be written as: %
\be T_{ab}^{[c]} = \frac i2 [C_2(so(c)_K), U^{(cc)}_{ab}]
+ i\sigma_c U^{(cc)}_{ab},\quad c=2,\dots ,n  \ee %
where $C_2(so(c)_K)$ is the second order Casimir of the $so(c)$
left action subalgebra, i.e.\ $C_2(so(c)_K)=\frac 12
\sum_{a,b=1}^c (K_{ab})^2$. The
generalized Gell-Mann formula can now be written as:%
\be T^{\sigma_2\dots \sigma_n}_{ab} = i\sum_{c = 2}^n \frac 12
[C_2(so(c)_K),
U^{(cc)}_{ab}]+  \sigma_c U^{(cc)}_{ab},\label{GGMv3}\ee %
which is to be compared with the original formula in the form
(\ref{Gell-Mann_original}). Again, the generalized formula
matches, by simplicity of the expression, the original one.
Besides, the very term when $c = n$ is, essentially, the original
Gell-Mann formula (since $C_2(so(n)_K)$ $=$ $C_2(so(n)_M)$),
whereas the rest of the terms can be seen as necessary corrections
securing the formula validity in the entire representation space.
The additional terms vanish for some representations yielding the
original formula.

The generalized Gell-Mann formula expression for the noncompact
``shear'' generators $T_{ab}$ holds for all cases of $sl(n,\R)$
irreducible representations, irrespective of their $so(n)$
subalgebra multiplicity (multiplicity free of the original
Gell-Mann formula, and nontrivial multiplicity) and whether they
are tensorial or spinorial. The price paid is that the Generalized
Gell-Mann formula is no longer solely a Lie algebra operator
expression, but an expression in terms of representation dependant
operators $K_{ab}$ and $U^{(cd)_{ab}}$.

\section{\label{sec:MatrixElements} Direct application - matrix elements of
$SL(n,\R)$ generators for all irreducible representations}

The generalized Gell-Mann formula, as given by (\ref{GGMv3}), can
be directly applied to yield all matrix elements of the
$\overline{SL}(n,\R)$ generators for all irreducible
representations, characterized by a complete set of labels
$\sigma_{i}$, $i=2,3,\dots , n$ (the invariant Casimir operators
are analytic functions of solely these labels), in the basis of
the maximal compact subgroup $Spin(n)$. Note that there can be
some additional discrete labels generally related to the finite
$\overline{SL}(n,\R)$ center group. Taking the matrix elements of
(\ref{GGMv3})
we get: %
{\renewcommand{\arraystretch}{0.2} \bea && \left<{
\begin{array}{l@{}l} \{J'\} & \\ \{k'\} & \{m'\} \end{array}}
\right| T^{\sigma_2\dots\sigma_n}_{ab} \left|
{\begin{array}{l@{}l} \{J\} & \\ \{k\} & \{m\} \end{array}}
\right>
                                                \label{Taction_cd} \\ %
&=& \left< { \begin{array}{l@{}l} \{J'\} & \\ \{k'\} & \{m'\}
\end{array}} \right| i\sum_{c = 2}^n \frac 12 [C_2(so(c)_K),
U^{(cc)}_{ab}] + \sigma_c U^{(cc)}_{ab}
 \left| {\begin{array}{l@{}l} \{J\} & \\ \{k\} & \{m\} \end{array}} \right>
                               \nonumber                       \\ %
&=& \frac i2 \sum_{c = 2}^n \big(C_2(so(c)_{ \{k'\} }) -
C_2(so(c)_{\{ k\} }) + \sigma_c \big) \left< {
\begin{array}{l@{}l} \{J'\} & \\ \{k'\} & \{m'\} \end{array}}
\right| U^{(cc)}_{ab} \left| {\begin{array}{l@{}l} \{J\} & \\
\{k\} & \{m\} \end{array}} \right>
                                                             \nonumber \\ %
&=& {\scriptstyle \frac i2\sqrt{\frac{dim(\{J\})}{dim(\{J'\})}}}
\sum_{c = 2}^n \big(C_2(so(c)_{\{ k'\} }) - C_2(so(c)_{\{ k\} }) +
\sigma_c \big)
 C\!\!{\scriptsize
\begin{array}{c@{}c@{}c} \{J\} & \boxbox & \{J'\} \\ \{k\} & (cc) & \{k'\}
\end{array}}
C\!\!{\scriptsize
\begin{array}{c@{}c@{}c} \{J\} & \boxbox & \{J'\} \\ \{m\} & (ab) & \{m'\}
\end{array}} \nonumber, \eea} %
where, in the last equality, the expression (\ref{Uaction}) for
the matrix elements of the $U$ operators is used. The second
Clebsch-Gordan coefficient, that is merely reflecting the
Wigner-Eckart theorem, can be evaluated in any suitable basis, not
necessarily the Cartesian one, due to the fact that the expression
is covariant with respect to the free index $(ab)$. Note, that
this is not the case for the first Clebsch-Gordan coefficient --
it is necessary in order to evaluate it to express the specific
vector ${\scriptsize \left|\begin{array}{c} \boxbox \\ (cc)
\end{array} \right>}$ in some basis that spans the entire vector
space over $Spin(n)$.

The final expression is simplified by choosing the indexes of the
generalized Gell-Mann formula matrix elements to be given by
labels of the $Spin(n) \supset Spin(n-1) \supset \cdots \supset
Spin(2)$ group chain representation labels. In this notation, the
basis vectors of the
$Spin(n)$ irreducible representations are written as: %
\be \left|\begin{array}{c} \{J\} \\ \{m\}
\end{array} \right> =
\left|\begin{array}{c} J_{Spin(n),1}\  J_{Spin(n),2} \ J_{Spin(n),3} \dots \\
J_{Spin(n-1),1}\  J_{Spin(n-1),2} \dots \\
\dots \\ J_{Spin(2)}
\end{array}  \right>. \ee%
Likewise, the set of indices $\{k\}$ of (\ref{naturalbasis}) is
thus given by the labels of the irreducible representations
$\{J_{Spin(n-1),1},$ $J_{Spin(n-1),2},$ $ \cdots;$
$J_{Spin(n-2),1}, J_{Spin(n-2),2},$ $ \cdots; $ $\dots;
J_{Spin(2)}\}$ of the $Spin(n) \supset Spin(n-1) \supset \cdots
\supset Spin(2)$ group chain.

To express the vector ${\scriptsize \left|\begin{array}{c} \boxbox \\
(cc)\end{array} \right>}$ in such a basis we notice first that it
corresponds to a diagonal traceless $n$ by $n$ matrix of the form
$diag(-\frac 1n,\dots , -\frac 1n$ $,\frac{n-1}{n},$ $-\frac 1n,
\dots ,-\frac 1n)$, with $\frac{n-1}{n}$ positioned at the $c$-th
row and column. On the other hand, the diagonal traceless matrix
$\sqrt{\frac 1{c(c-1)}} \,diag(-1,\dots, -1, c-1, 0,\dots ,0)$,
with first $c-1$ occurrences of $-1$, corresponds to a vector that
belongs to a second order symmetric tensor ($\boxbox$
representation) with respect to $Spin(c), Spin(c+1), \dots ,
Spin(n)$
subgroups, and it is invariant under $Spin(c-1)$:%
\be {\scriptsize
\left|\begin{array}{l} \{\boxbox\}_{Spin(n)}\\ \cdots \\
\{\boxbox\}_{Spin(c)}\\
\{0\}_{Spin(c-1)} \\ \cdots \\ 0
\end{array} \right>}\label{vect_cc}.
\ee%
This vector has $n-c+1$ double-boxes followed by $c-2$ zeros
underneath -- in shorthand notation: ${\scriptsize
\left|\begin{array}{l} \{\boxbox\}^{n-c+1} \\ \{0\}^{c-2}
\end{array} \right>}$. Somewhat peculiar is
the matrix $\sqrt{\frac 12} \,diag(-1,1, 0, 0,\dots)$ that corresponds to: %
\be {\scriptsize \left|\begin{array}{l} \{\boxbox\}^{n-1} \\
\{0\}^{0} \end{array} \right> \equiv {\textstyle
\frac{1}{\sqrt{2}}}
\left|\begin{array}{l} \{\boxbox\}^{Spin(n)}\\ \cdots \\
\{\boxbox\}^{Spin(4)}\\ 2 \\ 2
\end{array} \right>} + {\scriptsize {\textstyle \frac{1}{\sqrt{2}}}
\left|\begin{array}{l} \{\boxbox\}^{Spin(n)}\\ \cdots \\
\{\boxbox\}^{Spin(4)}\\ \ \ 2 \\ -2
\end{array} \right>}\label{vect_22},
 \ee%
where the standard labelling for $SO(n)$, $n \leq 3$ is implied,
in particular the $\boxbox$ representation corresponds to
$J_{Spin(3)} = 2$.

By combining these facts we find:%
\be {\scriptsize \left|\begin{array}{c} \boxbox \\
(cc)\end{array} \right> + {\textstyle \frac 1c} \sum_{d=c+1}^n
\left|\begin{array}{c} \boxbox \\
(dd)\end{array} \right> = {\textstyle
\sqrt{\frac{c-1}{c}}}\left|\begin{array}{l} \{\boxbox\}^{n-c+1} \\
\{0\}^{c-2} \end{array} \right>. }             \label{basis_connection}\ee%

However, when evaluating the $U^{(cc)}$ operators of (\ref{GGMv3})
in this basis, only the first term on the left-hand side is
relevant due to the fact
that:%
\be d>c \quad \Rightarrow \quad [C_2(so(c)_K), U^{(dd)}_{ab}] = 0.
\ee%

Having this in mind, we make use of (\ref{basis_connection}) to
recast, in the first equality of (\ref{Taction_cd}), the
$U^{(cc)}$ operators accordingly. Taking into account
arbitrariness of the $\sigma_c$ coefficients and following the
same steps as in (\ref{Taction_cd}), we finally obtain a rather
simple expression for the shear generator matrix elements for an
arbitrary $sl(n,\R)$ representation (labelled now by parameters
$\tilde
\sigma_c$): %
\be \begin{array}{c} \left< { \begin{array}{l@{}l} \{J'\} & \\
\{k'\} & \{m'\} \end{array}} \right| T_{\{w\}} \left|
{\begin{array}{l@{}l} \{J\} & \\ \{k\} & \{m\} \end{array}}
\right> = {\scriptstyle \frac
i2\sqrt{\frac{dim(\{J\})}{dim(\{J'\})}}}  C\!\!{\scriptsize
\begin{array}{c@{}c@{}c} \{J\} & \boxbox & \{J'\} \\ \{m\} & \{w\} & \{m'\}
\end{array}} \label{Taction}  \\
 \times \sum_{c =
2}^n {\textstyle \sqrt{\frac{c-1}{c}}} \Big( C_2(so(c)_{\{k'\}}) -
C_2(so(c)_{\{k\}}) + \tilde \sigma_c \Big)
 C\!\!{\scriptsize
\begin{array}{c@{}c@{}c} \{J\} & (\boxbox)^{n-c+1} & \{J'\} \\ \{k\} &
  (0)^{c-2} & \{k'\} \end{array}}
 .\end{array}\ee  %
The relation of the labelling of (\ref{Taction}) and the one of
(\ref{GGM}), i.e.\ (\ref{GGMv3}), is achieved provided $\sigma_c =
\tilde\sigma_c + \sum_{d=2}^{c-1} \tilde \sigma_d/d$. The
Clebsch-Gordan coefficient with indices $\{m\}, \{w\},  \{m'\}$ in
(\ref{Taction}) can be evaluated in an arbitrary basis (which is
stressed by denoting the appropriate index by $w$ instead by
$ab$). The other Clebsch-Gordan coefficient can be evaluated in
any basis labelled according to the $Spin(n) \supset Spin(n-1)
\supset \cdots \supset Spin(2)$ subgroup chain (e.g.
Gel'fand-Tsetlin basis) and can be, nowadays, rather easily
evaluated, at least numerically.

As already stated, the matrix elements of the $sl(n,\R)/so(n)$
operators, as given by the Generalized Gell-Mann formula, apply to
all tensorial, spinorial, unitary, nonunitary (both finite a
infinite-dimensional) $sl(n,\R)$ irreducible representations. In
many physics applications one is interested in the unitary
irreducible representations. The unitarity question goes beyond
the scope of the present work, and it relates to the Hilbert space
properties, i.e. the vector space scalar product. An efficient
method to study unitarity is to start with a Hilbert space
$L^2(Spin(n),\kappa)$ of square integrable functions with a scalar
product in terms of an arbitrary kernel $\kappa$, and to impose
the unitarity constraints both on the scalar products itself and
on the $sl(n,\R)/so(n)$ operators matrix elements in that scalar
product (cf. \cite{SijackiSL3}).

To sum up, the expressions (\ref{Maction}) and (\ref{Taction})
fully determine the action of the $sl(n,\R)$ operators for an
arbitrary irreducible representation given by the set of $n-1$
invariant Casimir operators labels $\tilde \sigma_c$. This action
is given in terms of the basis vectors (\ref{naturalbasis}) of the
representation spaces of the maximal compact subgroup $Spin(n)$ of
the $\overline{SL}(n,\R)$ group. This result is general due to a
Corollary of Harish-Chandra \cite{HarishChandra} that explicitly
applies to the case of the $sl(n,\R)$ algebras.

\section{Conclusion}

The Gell-Mann formula, as stated above, applies at the pure
algebraic level, i.e. as an algebraic expression, only in the case
of (pseudo) orthogonal algebras. One can formally write it in
other Lie algebra cases. However, it turns out, by an explicit
verification, that it is not generally valid, ie. that the closure
of the commutation relations of the generators given by this
formula is not granted generically. It turns out that certain
successful applications of the Gell-Mann formula, beyond the
orthogonal-like algebras, were actually carried out in a framework
of particular algebra representation spaces. Thus, the Gell-Mann
formula applicability can be broaden beyond the $so(m,n)$ algebra
cases, by utilizing it in other algebra cases provided some Lie
algebra representation conditions are met. As for the $sl(n,\R)$
algebras, contracted w.r.t. their $so(n)$ subalgebras, the
algebraic expression of the Gel-Mann formula matters generally for
the multiplicity free representations only. In a previous work,
starting from the known generic noncompact generators
representation expressions for $sl(3,\R)$ and $sl(4,\R)$, we found
the generalization of the Gell-Mann formula for the $sl(5,\R)$
case. It is valid for all representation spaces irrespective of
the $so(5)$ sub-representations non-trivial multiplicity. An
analysis of the structure of the generalized Gell-Mann formula in
the $sl(3,\R)$, $sl(4,\R)$ and $sl(5,\R)$ cases, especially of the
role played by the $K_{ab}$ operators that generate the
$SO(n)_{K}$, i.e. $Spin(n)_{K}$ group (acting to the left in the
group manifold and characterize the $SO(n)_{M}$, i.e.
$Spin(n)_{M}$ representations multiplicity) paved a way for a
successful generalization of the Gell-Mann formula for all
$sl(n,\R)$ algebras. The generalized formula is given by
expression (\ref{GGM}). It can take alternative forms, such as
(\ref{GGMv2}) and (\ref{GGMv3}), that suit better for certain
mathematical or physical applications. The generalized Gell-Mann
formula for the $sl(n,\R)$ and or $su(n)$ algebras, considered
w.r.t. their $so(n)$ subalgebras (the maximal compact subalgebra
of the $sl(n,\R)$ algebra) is compact and simple, and thus has a
great potential for both further general consideration and various
applications in mathematics and physics.

Note that the generalized Gell-Mann formula for the $sl(n,\R)$
algebras, that is valid for all representation Hilbert spaces, is
characterized precisely by a right number of $n-1$ (algebra rank)
parameters, i.e. representation labels. As a first and most
precious application, based on the generalized formula, we
obtained for the first time a closed form of the expressions of
all matrix elements of the $sl(n,\R)$ noncompact generators for
all irreducible representations. All representations meaning:
finite, infinite, tensorial and spinorial. A distinct feature of
our generalized Gell-Mann formula approach is that the generalized
expression goes beyond the standard notion of deformation of the
contracted algebra, as it depends on additional operators,
$K_{ab}$, not belonging, however directly related, to the
contracted algebra. Due to this fact, our generalization of the
Gell-Mann formula is remarkably simple (compared to complicated
polynomial expressions appearing in some other approaches to
generalize the Gell-Mann formula), nevertheless establishing a
direct relation between representations of the contracted and
original algebras.

\section{Acknowledgments}

This work was supported in part by MNTR, Project-141036. One of
us, I.S., would like to acknowledge hospitality and useful
discussions at the Institute for Nuclear Research and Nuclear
Energy in Sofia (Bulgaria) during his visit as early stage
researcher supported by the FP6 Marie Curie Research Training
Network "Forces-Universe" MRTN-CT-2004-005104.

\bibliographystyle{my-h-elsevier}

\end{document}